\documentclass[%
onecolumn,
superscriptaddress,
 amsmath,amssymb,
 aps, prl
]{revtex4-2}

\usepackage{graphicx}
\usepackage{dcolumn}
\usepackage{bm}
\usepackage{amsmath}
\usepackage[colorlinks =true,
            linkcolor = blue,
            urlcolor  = blue,
            citecolor = blue,
            anchorcolor = blue]{hyperref}
\usepackage{amssymb}
\usepackage{multirow}
\usepackage{braket}

\usepackage{xcolor}

\newcommand{\beq}{\begin{equation}}
\newcommand{\eeq}{\end{equation}}

\usepackage{mathtools}
\newcommand{\Avg}[1]{\ensuremath{\left\langle #1 \right\rangle}}

\usepackage{soul}
\usepackage{color}

\definecolor{orange}{rgb}{1.0, 0.5, 0.0}
\definecolor{violet}{rgb}{0.78,0.08, 0.52}
\definecolor{green}{rgb}{0.11, 0.35, 0.02}
\definecolor{bluebell}{rgb}{0.64, 0.64, 0.82}
\definecolor{capri}{rgb}{0.0, 0.45, 0.73}

\begin{document}

\preprint{APS/123-QED}

\title{Geometric phase amplification in a clock interferometer for enhanced metrology}  

\author{Zhifan Zhou}
\affiliation{Joint Quantum Institute, National Institute of Standards and Technology and the University of Maryland, College Park, Maryland 20742 USA}
\affiliation{Department of Physics, Ben-Gurion University of the Negev, Beer-Sheva 84105,
Israel}

\author{Sebastian C. Carrasco}
\affiliation{DEVCOM Army Research Laboratory, Adelphi, Maryland 20783, USA}

\author{Christian Sanner}
\affiliation{Department of Physics, Colorado State University, Fort Collins, Colorado 80523, USA}

\author{Vladimir S. Malinovsky}
\affiliation{DEVCOM Army Research Laboratory, Adelphi, Maryland 20783, USA}

\author{Ron Folman}%
\affiliation{Department of Physics, Ben-Gurion University of the Negev, Beer-Sheva 84105,
Israel}

\date{\today}%

\begin{abstract} 
High-precision measurements are crucial for testing the fundamental laws of nature and for advancing the technological frontier. Clock interferometry, where particles with an internal clock are coherently split and recombined along two spatial paths, has sparked significant interest due to its fundamental implications, especially at the intersection of quantum mechanics and general relativity. Here, we demonstrate that a clock interferometer provides metrological improvement with respect to its technical-noise-limited counterpart employing a single internal quantum state. This enhancement around a critical working point can be interpreted as a geometric-phase-induced signal-to-noise ratio gain. In our experimental setup, we infer a precision enhancement of 8.8 decibels when measuring a small difference between external fields. We estimate that tens of decibels of precision enhancement could be attained for measurements with a higher atom flux. This opens the door to the development of a superior probe for fundamental physics as well as a high-performance sensor for various technological applications. 
\end{abstract}

\maketitle

Matter--wave interferometry has evolved into a robust platform for probing the fundamental principles of quantum mechanics\,\cite{cronin2009optics, KasevichS2002}. It is instrumental in sensing gravitational signals\,\cite{cronin2009optics}, testing the quantum superposition principle\,\cite{kovachy2015quantum}, and in exploring the intersection between quantum mechanics and general relativity\,\cite{Zych11, sinha2011atom, ClockScience, Pikovski2015, NJP2017, QuCom, Giese21}. Furthermore, matter--wave interferometry is a fundamental demonstration of our ability to manipulate atomic motion coherently\,\cite{KasevichS2002, KochEPJQT2022, DashAVSQS2024}. A typical matter--wave interferometry experiment involves the creation of spatially superposed atomic wave packets, acquisition of a relative phase signal between them, subsequent interference of the wave packets, and some form of measurement of the final signal. Naturally, these operations contribute to the experimental noise budget, pushing the sensitivity away from the standard quantum limit\,(SQL), the lower sensitivity bound of metrology without quantum entanglement. Here, we use the internal structure of two-level atom matter waves and a steep change in the geometric phase\,(GP) response (rooted in the Pancharatnam--Berry phase\,\cite{berry1984quantal, pancharatnam1956generalized, GeoPhaseJump}) to amplify a small interferometric signal. The primary advantage of the scheme appears in the observation that the technical noise is not amplified with the interferometric signal, thus increasing the overall signal-to-noise ratio. 
Moreover, our approach is an example of clock interferometry that exploits an interplay of internal (clock) and external (spatial) degrees of freedom\,\cite{ClockScience, QuCom}. Our study is a proof-of-principle experiment highlighting the potential of geometric phase amplification for a wide range of precision measurements.

\begin{figure*}[h!]
\centering
\includegraphics[scale=0.52]{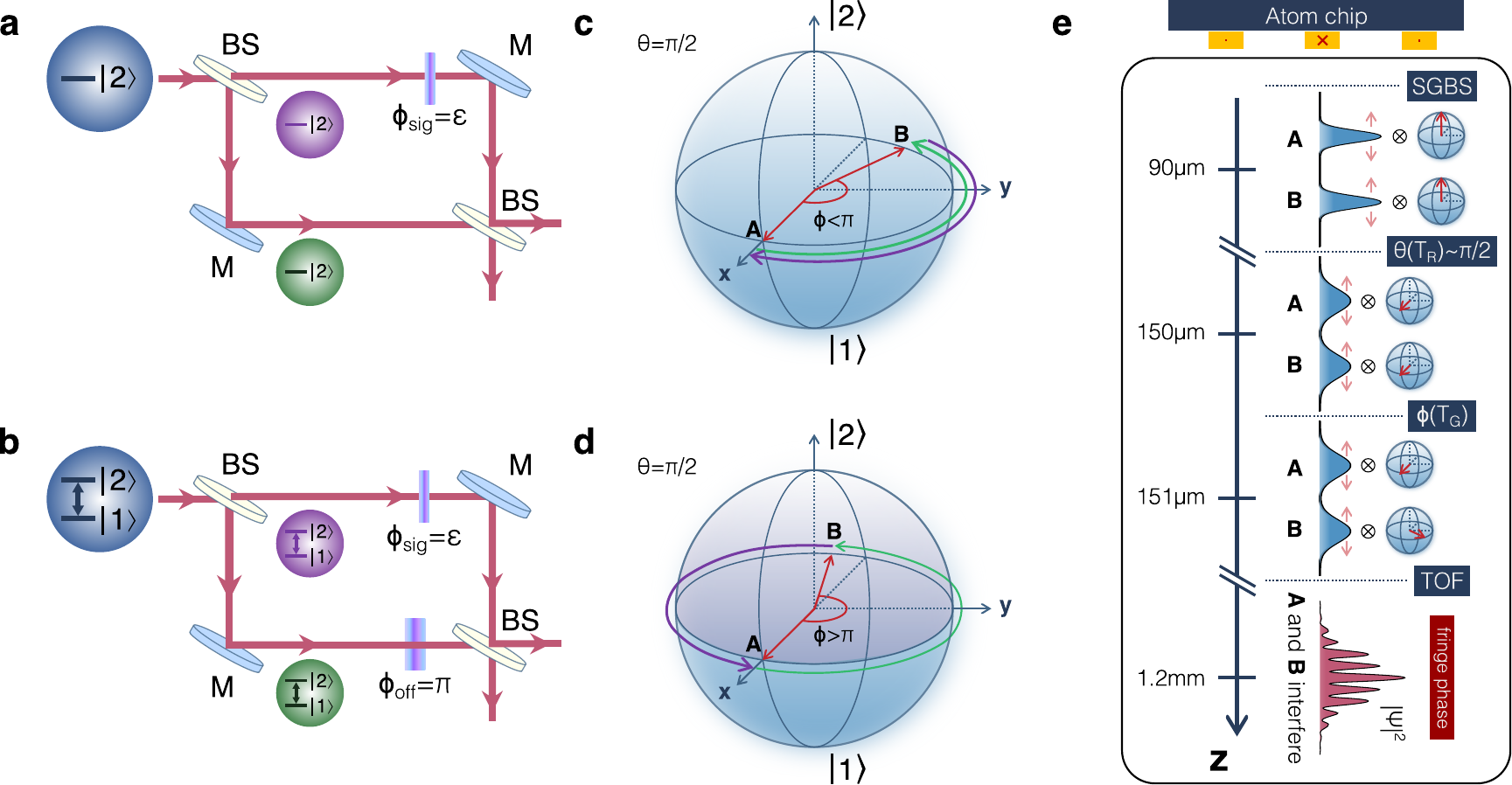}
\caption{\label{fig-Scheme} \textbf{Illustration of the Geometric Phase Amplification in a Matter--Wave Interferometer and the Experimental Sequence.} In \textbf{a}, we show the case in which there is the same single state in each arm, where $\phi_\text{sig}$ is a weak interferometric signal due to the coupling to the external field. In \textbf{b}, we present the case in which the matter wave is prepared in an internal superposition of the states $\ket{1}$ and $\ket{2}$. In \textbf{c}, we illustrate that when the relative rotation of the Bloch vector is less than $\pi$, the geodesic connection line matches the latitude line, yielding no enclosed areas and thus a geometric phase of zero. Conversely, when the relative rotation of the Bloch vector surpasses $\pi$\,(illustrated in \textbf{d}), the geodesic and latitude lines encircle the north hemisphere, leading to a geometric phase shift of $\pi$ when $\theta$ is $\pi/2$. In \textbf{e}, we display the experimental realisation of the sequence for the longitudinal interferometer\,(illustration not to scale). The experiment unfolds in free fall along the z-axis, representing gravity's direction. Initially, we achieve a coherent spatial splitting using a Stern--Gerlach beam splitter\,(SGBS)\,\cite{machluf2013coherent} combined with a stopping pulse. This creates two wave packets in the $\ket{2}$ state. We then initialise the internal superposition using a radio frequency\,(RF) pulse lasting $T_R$. Following this, we can encode a relative phase of the two wave packets by applying a magnetic field gradient $\partial B/\partial z$ for a duration of $T_G$. Three parallel gold wires on the chip surface, with the same current flowing through them in alternating directions, produce the magnetic gradient pulses. The final interference pattern, from which we derive the interference phase, forms during the time-of-flight\,(TOF) free evolution as the wave packets expand and overlap. For further details regarding the experiment, see SM.}
\label{fig1}
\end{figure*}

We start by describing the matter--wave interferometry scheme we use. First, we produce two wave packets $\bf A$ and $\bf B$ spatially superposed and in the same internal two-level state. As the wave packets evolve in an inhomogeneous external field, each internal degree of freedom accumulates a differential phase because of the different coupling with the field. For notational simplicity, we map the relative phase accumulation into wave packet \textbf{B}. We can then describe the wave packets by the following wavefunctions:
\begin{equation}
\Psi_\textbf{A} = \psi_\textbf{A}(\vec r) \ket{a} = \psi_\textbf{A}(\vec r) \left(\cos \frac{\theta}{2} \ket{2} + \sin \frac {\theta}{2} \ket{1}\right)\, ,~~\\
\Psi_\textbf{B} = \psi_\textbf{B}(\vec r) \ket{b} = \psi_\textbf{B}(\vec r)  \left(e^{i\phi_2}\cos \frac{\theta}{2} \ket{2} + e^{i\phi_1} \sin \frac {\theta}{2} \ket{1}\right)\, ,
\label{EQ-states}
\end{equation}
where $\psi_\textbf{A}(\vec r)$ and $\psi_\textbf{B}(\vec r)$ are the spatial components of the wavefunctions, $\ket{1}$ and $\ket{2}$ are two internal degrees of freedom that form the internal superposition, and $\phi_1$ and $\phi_2$ are the accumulated phases. The interference phase is given by 
\begin{equation}
\Phi_T=\arg\ \Avg{\Psi_\textbf{A} | \Psi_\textbf{B}} =  \phi_2 + \arctan \left \{ \frac{\sin^2(\theta/2)\sin(\phi_1 - \phi_2)}{\cos^2(\theta/2)+\sin^2(\theta/2)\cos(\phi_1 - \phi_2)} \right \}\, . 
\label{EQ-Phi}
\end{equation}
Thus, the phases $\phi_{1}$ and $\phi_{2}$ lead to a phase shift in the interference pattern. If $\theta=0$, the entire population is in state $\ket{2}$, and $\Phi_T = \phi_2$. Similarly, if $\theta=\pi$, the entire population is in state $\ket{1}$, and $\Phi_T = \phi_1$. In the limit  $\phi = \phi_2 - \phi_1 = \pi - \varepsilon$ with $\varepsilon \ll 1$, we obtain an approximated linear dependence given by
\begin{equation}
\Phi_T= \phi_2 - G  \varepsilon \,
\label{EQ-Phi-app}
\end{equation}
where $G =\,(1 - R)^{-1}$ is the slope, and $R = \cos^2(\theta/2) / \sin^2(\theta/2)$ is the ratio between the populations in the internal degrees of freedom. In the regime where $R$ is approximately 1, the signal is boosted by $G$, and the interference visibility $v$ approaches zero\,(see more details in supplementary material\,[SM] and \cite{machluf2013coherent} for the definition of visibility $v$), as the following relationship is fulfilled: $v^2=1-4\sin^2(\theta/2)\cos^2(\theta/2)\sin^2(\phi/2)$\,\cite{NJP2017,QuCom}. To drive the system to the working point\,($\phi \approx \pi$), the introduction of a $\pi$ phase offset is needed. To do that, one can use additional manipulations, such as the coherent coupling to additional levels discussed in\,\cite{jaksch2000fast}. This positions the system at the optimal sensitivity point for practical operation. Fig.\,\ref{fig-Scheme}\textbf{a} and Fig.\,\ref{fig-Scheme}\textbf{b} show the interferometric scheme for two different cases. Fig.\,\ref{fig-Scheme}\textbf{a} shows the case where $\theta = 0$, while Fig.\,\ref{fig-Scheme}\textbf{b} illustrates the case where $\theta$ is close to $\pi/2$; the phase is biased at $\pi$, and there is phase amplification. 

At first glance, one may argue that interferometry in the low-visibility region presented in Fig.\,\ref{fig-Scheme}\textbf{b} would be impractical as the phase uncertainty diverges for $v \to 0$. As we show in the following, this region presents metrological improvement compared to $\theta = 0$ or $\theta=\pi$, where the internal degrees of freedom do not play an active role, as only one internal state is populated. Two conditions have to be met to obtain metrological improvement. First, $\phi$ has to be close to $\pi$, which is the working point of the interferometer. Second, $\theta$ must be close enough to $\pi/2$ to increase the signal but not precisely $\pi/2$, where the visibility is zero. Once these two conditions are satisfied, the phase uncertainty will not diverge, and the phase response will be amplified [as shown in Eq\,\eqref{EQ-Phi-app}] without a proportional rise in technical noise. This can effectively suppress the negative effect of the technical noise on the interferometric precision. Typical matter--wave interferometers suffer from technical noise due to the initial preparation, path-length noise rooted in driving field noise, detection noise due to the limited resolution of the camera, and environmental noise, among others\,\cite{cronin2009optics}. This technical noise becomes more important when the number of atoms increases, as the quantum noise decreases\,\cite{HighFlux}. 

Fig.\,\ref{fig-Scheme}\textbf{e} depicts the experimental scheme\,(see more detail in \cite{QuCom, GeoPhaseJump}). First, we prepare a Bose-Einstein condensate\,(BEC) with rubidium-87 atoms in the internal state $\ket{2}\equiv\ket{F=2, m_F=2}$, where $F$ represents the total angular momentum and $m_F$ its magnetic projection. After that, the wave packet is split into two using a Stern--Gerlach-type beam splitter technique\,\cite{machluf2013coherent}. We then apply a uniform radio-frequency\,(RF) pulse to drive the population into state $\ket{1\rangle\equiv|F=2,m_F=1}$. Now, the internal states of both wave packets are a superposition of $\ket{2}$ and $\ket{1}$ with a polar angle $\theta$\,(point $\bf A$ on the Bloch spheres in Fig.\,\ref{fig-Scheme}\textbf{c}--\ref{fig-Scheme}\textbf{d}). We next apply a magnetic-field gradient; thus, both wave packets evolve under different magnetic fields. Due to the differing magnetic projections of $\ket{1}$ and $\ket{2}$, the wave packet rotates in the Bloch spheres until it points at $\bf B$ on the Bloch spheres\,(Fig.\,\ref{fig-Scheme}\textbf{c}--\ref{fig-Scheme}\textbf{d}). Now, we have the previously discussed wavefunctions\,[see Eq.\,\eqref{EQ-states}] with $\phi_2 = 2 \phi_1$. Finally, the wave packets are overlapped via time-of-flight expansion, and the visibility $v$ and phase $\Phi_T$ can be extracted by fitting the interference pattern (see SM and \cite{machluf2013coherent}). If $\theta$ is close enough to $\pi/2$, we expect a phase amplification effect as Eq.\,\eqref{EQ-Phi-app} predicts.

\begin{figure*}[h!]
\centering
\includegraphics[scale=0.5]{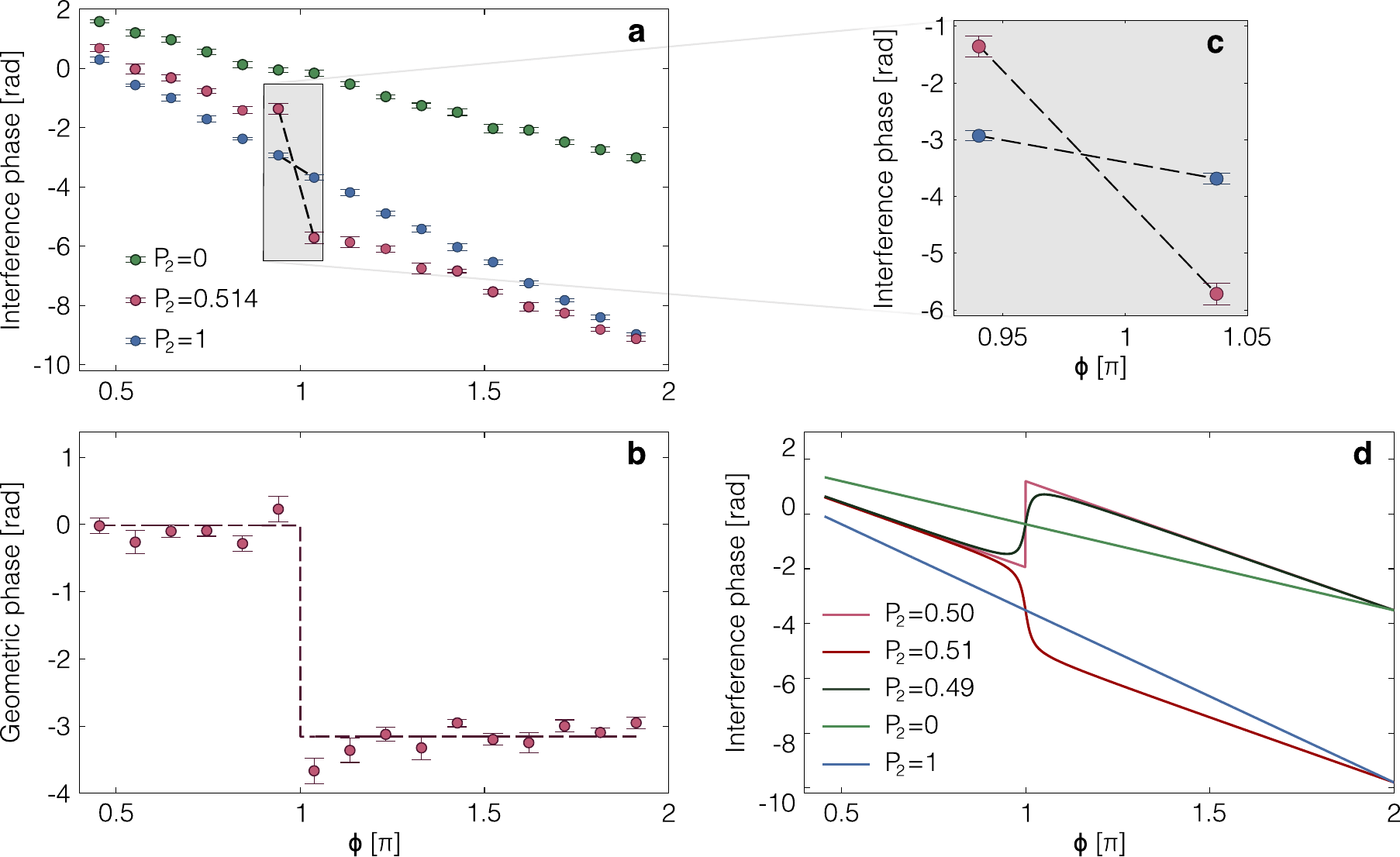}
\caption{\textbf{Experimental Results of the Geometric Phase Enhancement.} In \textbf{a}, we measure the interference phase\,(red circles) when each wave packet is prepared with 51.4$\%$ and 48.6$\%$\,(and 0.4$\%$ uncertainty) of the population in state $\ket{2}$ and $\ket{1}$, respectively. Phase measurements using single-state interference to demonstrate linear phase accumulation are shown in green and blue circles. Each data point is an average of eight experimental cycles, and the error bars are the standard error
of the mean (SEM) in this subsample. In \textbf{b}, we remove the dynamical phase\,(estimated analytically from the population distribution in state $\ket{2}$) to illustrate the geometric nature of the phase amplification. We compute the dashed purple curve using $\Phi_G = \Phi_T - \Phi_D$, according to the analytical expressions in the main text. In \textbf{a} and \textbf{b}, we choose the appropriate reference phase so that the GP starts at zero, thus matching the geodesic rule. In \textbf{c}, we gather the two pairs of data points that will be used to calculate the corresponding sensitivities. The raw data for four single shots, representing these four data points, is shown in Fig.\,\ref{SingleShot}. We connect each pair of points with a black dashed line as in \textbf{a}. We obtain 8.8 dB of enhancement in our $\Delta \phi^2$ sensitivity estimation. Note that the enhancement comes primarily from the phase amplification that overcomes the minor uncertainty increase\,(see discussion that accompanies Fig.\,\ref{PhaseErrVsVisibility}). In \textbf{d}, we display model traces illustrating the transition from single-state to two-state interference, exhibiting the phase amplification. The raw data presented here were gathered at the Atom Chip Lab at Ben-Gurion University of the Negev\,\cite{QuCom, GeoPhaseJump}.}
\label{fig2}
\end{figure*}

We can understand the phase-shift amplification in a GP picture via the geodesic rule proposition\,\cite{samuel-bhandari, bhandari19912, GeoPhaseJump}. The measured phase has two components. One part is the GP, which results from non-cycling evolution as in Fig.\,\ref{fig-Scheme}\textbf{c}--\ref{fig-Scheme}\textbf{d}, and the other is a dynamical phase\,(DP), a linear component resulting from the magnetic projection. To be precise, $\Phi_T = \Phi_G + \Phi_D$, where the first term is the GP, and the other, the DP, is given by $\Phi_D = \phi\,(1 - \cos \theta) / 2$. The GP follows the geodesic rule that dictates the reverse progression along the geodesic curve, which is the shortest path on the surface of the Bloch or Poincaré sphere, linking the initial and final points. The GP is half the area enclosed by the trajectory and the corresponding geodesic path. For instance, when the internal states of $\bf A$ and $\bf B$ are in an equal superposition, the Bloch vectors are on the equator of the Bloch sphere. If the relative rotation or azimuth angle between $\bf A$ and $\bf B$ is less than $\pi$, the geodesic and latitude lines coincide; thus, they do not enclose any actual area\,(as shown in Fig.\,\ref{fig-Scheme}\textbf{c}). In contrast, when this rotation exceeds $\pi$, they collectively encompass a hemisphere\,(Fig.\,\ref{fig-Scheme}\textbf{d}), resulting in a geometric $\pi$ phase change. If $\theta$ is not equal to $\pi/2$ but close, there is no sudden phase change but a region with an increased slope. This geometric $\pi$ phase change has been experimentally demonstrated in recent studies\,\cite{GeoPhaseJump} and is related to the Pancharatnam phase\,\cite{pancharatnam1956generalized}. Nevertheless, the metrological properties have never been previously analyzed in detail. Notably, the GP depends only on the parameter-space area. Thus, our scheme is, by design, resilient to certain disturbances and imperfections. This GP resilience has been both theoretically and experimentally studied in spin-half systems with cyclic evolutions\,\cite{Vedral2003, GP2003, leek2007observation, Rauch2009, BerryExperi2013, yale2016optical}. In the broader context of non-cyclic evolution, we observed the rapid phase change along the geodesic rule, amplification of a small phase change signal, and lack of amplification of technical noise. Thus, the hindering effect of technical noise on interferometric precision is expected to be suppressed.

Fig.\,\ref{fig2} depicts the main experimental results\,(the raw data presented here were gathered at the Atom Chip Lab at Ben-Gurion University of the Negev\,\cite{QuCom, GeoPhaseJump}). Each data point is an average of eight experimental cycles. The small data sample shows that the method works well without requiring prohibitively large data samples. In fact, as can be seen from the fit errors in Fig.\,\ref{SingleShot}, for some applications, single shots would be enough. In Fig.\,\ref{fig2}\textbf{a}, the red circles show a rapid phase change when 51.4$\%$ and 48.6$\%$ of the population\,(with a 0.4$\%$ standard error of the mean\,[SEM]) is in state $\ket{2}$ and $\ket{1}$, respectively. The population of 51.4$\%$ is about three SEM and one standard deviation away from an equal partition. Here, $\phi$ is scanned around the working point $\phi = \pi$, as suggested by the theoretical model. In contrast, the green and blue circles illustrate the case where the population is either in state $\ket{1}$ or $\ket{2}$. Thus, a conventional phase measurement from a single-state interference follows a linear phase accumulation process without amplification. Fig.\,\ref{fig2}\textbf{a} also displays the SEM of the measured interferometric phase, which slightly increases around the working point. Taking the increase in slope and noise into account, we estimate within the steep-slope region a measurement precision improvement of 8.8 dB with respect to the best conventional counterpart\,(interferometry with $\ket{2}$) experiencing the same noise floor. The direct comparison of the four data points connected by black dashed lines in Fig.\,\ref{fig2}\textbf{a} is highlighted in Fig.\,\ref{fig2}\textbf{c}\,(see the detailed sensitivity calculation in SM). Fig.\,\ref{fig2}\textbf{b} illustrates the geometric character of the phase amplification by removing the DP to isolate the GP. The phase remains rigid until it changes rapidly around $\phi = \pi$. Fig.\,\ref{fig2}\textbf{d} presents the analytical results for the interference phase across various population distributions, predicting the phase change depicted in Fig.\,\ref{fig2}\textbf{a}. 

\begin{figure*}
\centering
\includegraphics[scale=0.5]{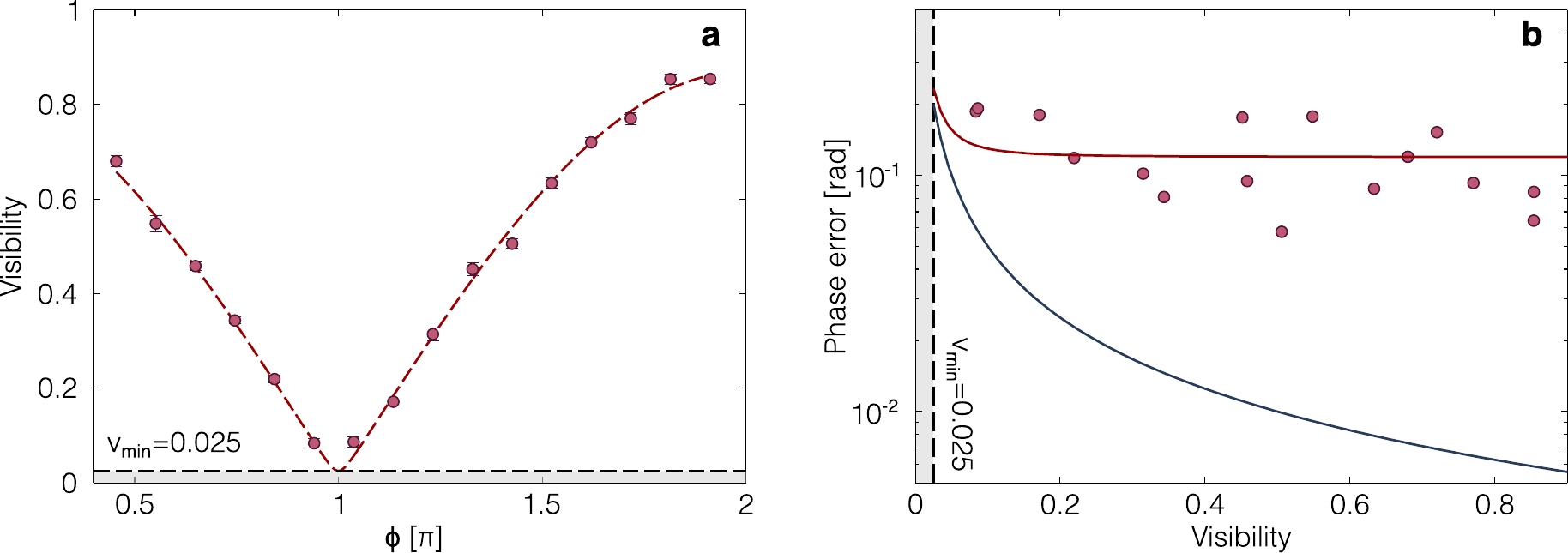}
\caption{\textbf{Visibility and Measurement Uncertainty.} In \textbf{a}, we display the visibility of the interference pattern as a function of the relative rotation. The populations in the $\ket{2}$ and $\ket{1}$ states are measured as 51.4$\%$ and 48.6$\%$. The states are not equally populated, and therefore, the visibility does not reach zero. Instead, the minimal visibility is 0.025\,(see main text). Each data point is an average of eight experimental cycles, and the error bars are the SEM in this subsample. In \textbf{b}, we present how the phase error\,(SEM) varies as a function of visibility. The data suggest that the measured phase fluctuations do not change over a wide visibility range, which is expected when technical noise is the dominant contribution. The red line is derived from a straightforward model where a constant technical noise of approximately 0.1 rad is added to a quantum noise component\,(blue line), which varies as $(v \, \sqrt{N} \, \sqrt{A})^{-1}$, where $v$ is the visibility, $N$ is the atom number of $5\times10^3$, and $A$ is the number of experimental cycles\,(eight). Note that the phase error of the data reaches 0.2\,rad at low visibility, consistent with the $\pm$\,0.2\,rad presented in Fig.\,\ref{fig2}\textbf{c}. The raw data presented here were gathered at the Atom Chip Lab at Ben-Gurion University of the Negev\,\cite{QuCom, GeoPhaseJump}.}
\label{PhaseErrVsVisibility}
\end{figure*}

To better understand the reason for the metrological improvement, we explain how the phase measurement noise depends on $\phi$ in Fig.\,\ref{fig2}\textbf{a}. When $\phi$ approaches $\pi$ (see Fig.\,\ref{PhaseErrVsVisibility}\textbf{a}), the visibility slowly diminishes in accordance with theory, $v^2=1-4\sin^2(\theta/2)\cos^2(\theta/2)\sin^2(\phi/2)$\,\cite{NJP2017, QuCom}. In an interferometer only limited by quantum noise, the theory predicts an increase in phase measurement noise proportional to $1/v$ as extracting the phase with low visibility becomes more difficult. However, practical matter--wave interferometer implementations suffer from additional technical noise sources. Fig.\,\ref{PhaseErrVsVisibility}\textbf{b} plots measured phase error as a function of visibility\,(red circles) together with two theoretical models. The blue solid line shows the $1/v$ scaling in the phase error that one would expect for an interferometer operating at the SQL. The dashed red line shows the same model with the addition of technical noise of 0.1 rad. This modification to the $1/v$ model agrees well with the data points that show the dependence of the phase measurement uncertainty with respect to visibility\,(see Fig.\,\ref{PhaseErrVsVisibility}\textbf{b}). Note that as $\theta \neq \pi/2$, the diverging part of Fig.\,\ref{PhaseErrVsVisibility}\textbf{b} is unreachable as there is a minimum visibility value\,(highlighted in both panels).

Accordingly, the uncertainty for the phase measurement has two components: fundamental atom shot noise or quantum noise that scales with the reduction in visibility, and technical noise, $\Delta \Phi_T^2 = \Delta \Phi_\text{quantum}^2 + \Delta \Phi_\text{technical}^2$. The measurement uncertainty for the phase $\phi$, which is the minimal phase that can be resolved, is given by
\begin{equation}
\Delta \phi = \frac{\Delta \Phi_T}{|\partial \Phi_T / \partial \phi|}\, .
\end{equation}
For large enough values of $G = \partial \Phi_T / \partial \phi$, the technical noise does not contribute significantly to the sensitivity and $\Delta \phi \approx \Delta \Phi_\text{quantum}/G$. In that case, only quantum noise limits the sensitivity. In contrast, when employing one internal state, the sensitivity is limited by the technical noise. 

\begin{figure*}
\begin{center}
\includegraphics[scale=0.5]{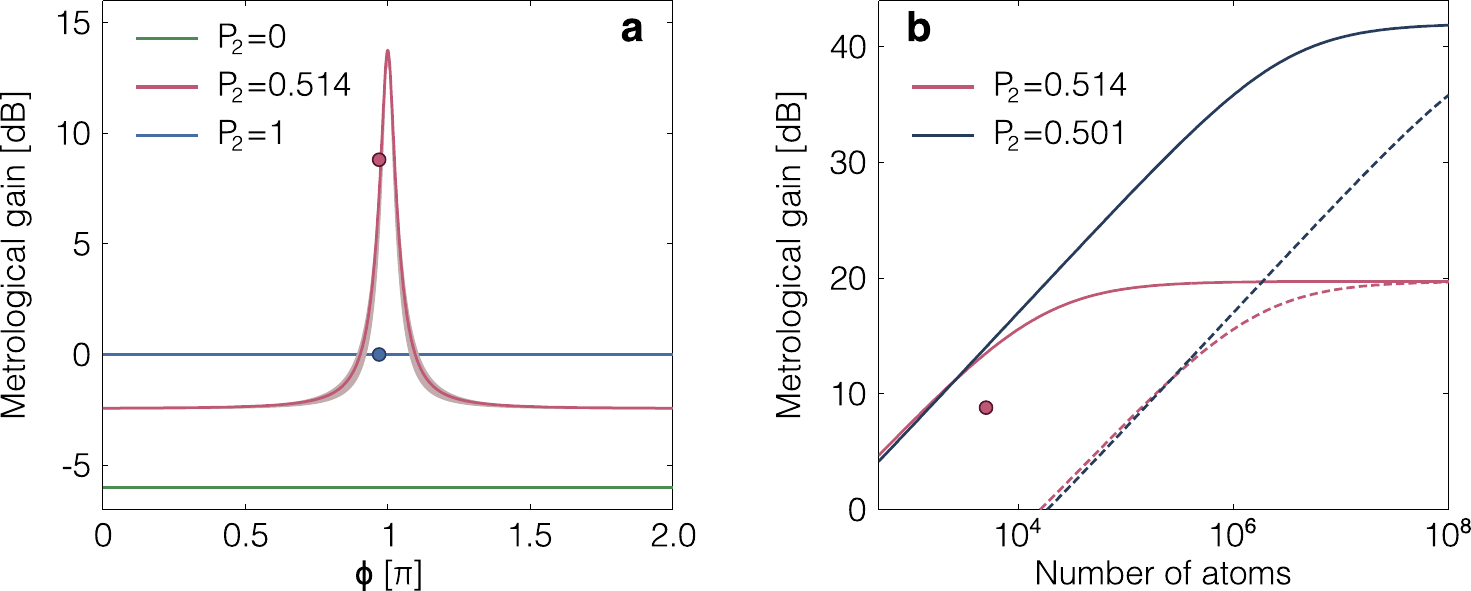}
\end{center}
\caption{\textbf{Metrological Gain with Different Experimental Parameters.} In \textbf{a}, we present the estimated metrological gain with respect to the standard single-state interferometer\,(the entire population in state $\ket{2}$ or $\ket{1}$) versus the relative rotation $\phi$, for selected values of the population in states $\ket{1}$ and $\ket{2}$. The red point represents the experimental measurement of $8.8$ dB with the internal superposition\,($P_2=0.514$), and the number of atoms is $5\times10^3$, and the blue point represents the reference with a single-spin state\,($P_2=1$). As the magnetic projection of state $\ket{2}$ is twice that of state $\ket{1}$, there is 6 dB of difference, assuming the noise to be the same. The red solid curve indicates a situation with an internal superposition of $P_2 = 0.514$, while the gray area represents the results considering a 0.4$\%$ uncertainty in the internal populations. Another source of uncertainty are the fluctuations in the magnetic gradient that generates the phase $\phi$, which are estimated to give rise to a $2.2\%$ uncertainty in $\phi$\,(smaller than the width of the data points). This is consistent with previous calculations\,(specifically in \cite{machluf2013coherent, margalit2019analysis}). We offer more details in the noise analysis section of the SM. In \textbf{b}, we show the estimated metrological gain as a function of the number of atoms $N$. The blue solid curve has the internal superposition of $P_2 = 0.501$, which generates higher metrological gain with a higher phase-change slope, as pointed out in Eq.\,\eqref{EQ-Phi}. Note that the experimental point is lower than the theoretical curve as it was not taken exactly at $\phi=\pi$, while the curves consider the optimal case of $\phi=\pi$. All solid curves involve an estimated technical noise of $0.1$ rad, as consistent with our described experimental conditions\,(see more details in SM). Regarding metrology, a technical noise of $0.1$ rad is realistic for a miniaturised apparatus on a chip\,\cite{keil2016fifteen, amico2021roadmap}. Dashed curves show the prediction for a technical noise of $0.01$ rad. We use the sensitivity of the case $P_2=1$ as a reference to define the metrological gain.}
\label{EstimatedGain}
\end{figure*}

Fig.\,\ref{EstimatedGain}\textbf{a} shows how the estimated metrological gain depends on the relative rotation $\phi$ for selected values of the population in states $\ket{1}$ and $\ket{2}$. To calculate this, we compute the phase shift uncertainty as a function of $\phi$ using the model $\Delta \Phi_T^2 =\,(v \sqrt{N A})^{-2} + \Phi_\text{technical}^2$, which agrees with the experimental data previously discussed. In addition, we evaluate the slope $G = \partial\Phi_T/\partial \phi$ from Eq.\,\eqref{EQ-Phi}. As a reference to define the metrological gain, we use the measurement uncertainty for the phase $\phi$ obtained when operating the interferometer at $P_2 = 1$ (the black pentagon in Fig.\,\ref{EstimatedGain}\textbf{a}). We estimate a metrological gain of approximately 10 dB around the working point $\phi = \pi$ when $P_2 = 0.514$. That value is close to the 8.8 dB extracted from the experimental results (the red star in Fig.\,\ref{EstimatedGain}\textbf{a}). We attribute the difference to an underestimation of the slope due to the lack of sampling resolution in the vicinity of the working point. For reference, we also include a plot of the expected results using a single quantum state, for $P_2 = 0$ and $P_2 = 1$, which differ due to the difference in the magnetic projection between states.

In the low atom flux limit, the quantum noise is relatively large\,\cite{anders2021momentum}. This limits the metrological gain that can be attained using geometric phase amplification. We anticipate that a high-flux atom interferometer\,\cite{HighFlux} will alleviate this limit and increase the impact of the geometric phase amplification. In Fig.\,\ref{EstimatedGain}\textbf{b}, we show how the metrological gain increases with $N$. Furthermore, if $N$ is large and the population ratio is closer to one\,(see the $P_2 = 0.501$ case), the metrological gain could be even higher\,(up to 40 dB in the blue solid curve) due to a steeper phase response. We assume that the technical noise is the same\,(0.1 rad) when calculating these predictions. In Fig.\,\ref{EstimatedGain}\textbf{b}, we also include dashed black and red curves showing the prediction for a technical noise of 0.01 rad as an example of what happens with smaller values of technical noise. In this case, our technique still offers metrological improvement. However, the atom flux threshold upon which we start to observe it increases. 

As an outlook, let us emphasise that this work should be regarded as a proof-of-principle experiment highlighting opportunities for progress in sensors based on the non-cyclic evolution adhering to the geodesic rule. An additional advantage not discussed here is the possible fast phase accumulation due to the possibility of a non-adiabatic GP accumulation\,\cite{sjoqvist2012non, xu2019breaking, PhysRevLett.127.030502}. Further development is required to adapt this method to different physical observables that one would like to measure, where any observable that changes the ``tick rate" of the clock by a differential interaction with its two levels may be detected, e.g., DC and AC Stark shifts, if there exists a gradient across the interferometer. For example, one may envision the magnetic gradient used in this work as analogous\,\cite{ClockScience, QuCom} to a difference in proper time that a gravitational field would produce and our technique as comparing the tick rates of a quantum clock (represented by the internal states) in a spatial superposition (probing two regions of space). Therefore, the experimental scheme could work as a gravitational sensor if high-precision clocks able to resolve the redshift are used\,\cite{PRXQuantum}. Thus, the technique introduced here may be suitable for studying the interplay between general relativity and quantum mechanics (and potentially testing exotic concepts such as the discreteness of time\,\cite{christodoulou2022experiment}). Finally, the method described here may be combined with further quantum noise reduction technologies, such as spin squeezing\,\cite{carrasco2022extreme}. 

To conclude, we have demonstrated that atom interferometry employing matter waves with internal degrees of freedom\,(so-called clock interferometry) can be metrologically advantageous. This improvement is due to the geometric phase amplification that arises from non-cyclic evolution, which results in a steep phase response that is not accompanied by a proportional rise in noise. We experimentally demonstrated 8.8 dB of metrological improvement with respect to the case without geometric phase amplification. We estimate that tens of decibels of enhancement could be attained by increasing the atom flux and adjusting the ratio between the internal states. Since long lifetime clock transitions have been employed for matter--wave interferometry\,\cite{Tino2017}, we anticipate our scheme would offer significant advancements in pushing the limits of matter--wave sensing capabilities\,\cite{kovachy2015quantum, overstreet2022observation}. In summary, this work offers insights into the potential of the geometric phase amplification to boost metrological sensitivity and may be implemented in numerous applications. 

\section*{Acknowledgments}
We thank the BGU nano-fabrication facility for the high-quality chip and the BGU experimental team for their support in acquiring the data. Funding: This work was funded, in part, by the Israel Science Foundation\,(grants no. 856/18, 1314/19, 3515/20, and 3470/21), the German-Israeli DIP project\,(Hybrid devices: FO 703/2-1) supported by the DFG, and the European ROCIT consortium for stable optical clocks. This research was also supported by the Army Research Laboratory under Cooperative Agreement Number W911NF-21-2-0037\,(SC). We thank Yair Margalit, Daniel Rohrlich, Yonathan Japha, Samuel Moukouri, and Yigal Meir for their helpful discussions.

\section*{Author Information}
{\small Correspondence and requests for materials should be addressed to Sebastian C. Carrasco (\url{seba.carrasco.m@gmail.com}), 
Christian Sanner (\url{sanner@colostate.edu}), and Ron Folman.(\url{folman@bgu.ac.il}).}

\bibliography{sample}
\newpage
\newpage
\newpage
\newpage

\pagebreak

\pagebreak

\pagebreak

\bigskip
\bigskip
\bigskip
\newpage

\newpage

\centerline{Supplementary material for}
\bigskip
\bigskip

\centerline{\bf Geometric phase amplification in a clock interferometer for enhanced metrology}

\bigskip

\centerline{Zhifan Zhou$^{1,2}$, Sebastian C. Carrasco$^{3}$, Christian Sanner$^{4}$, Vladimir S. Malinovsky$^{3}$, and Ron Folman$^{2}$}

\centerline{$^{1}$\it Joint Quantum Institute, National Institute of Standards and Technology}

\centerline{\it and the University of Maryland, College Park, Maryland 20742 USA}

\centerline{$^{2}$\it Department of Physics, Ben-Gurion University of the Negev, Beer-Sheva 84105, Israel}

\centerline{$^{3}$\it DEVCOM Army Research Laboratory, Adelphi, Maryland 20783, USA}

\centerline{$^{4}$\it Department of Physics, Colorado State University, Fort Collins, Colorado 80523, USA}

\bigskip

\bigskip

\renewcommand{\theequation}{S\arabic{equation}}
\renewcommand{\thefigure}{S\arabic{figure}}
\setcounter{figure}{0}
\setcounter{equation}{0}

\section{Detailed experimental procedure}
\begin{figure*}
\centering
\includegraphics[width=0.9\textwidth]{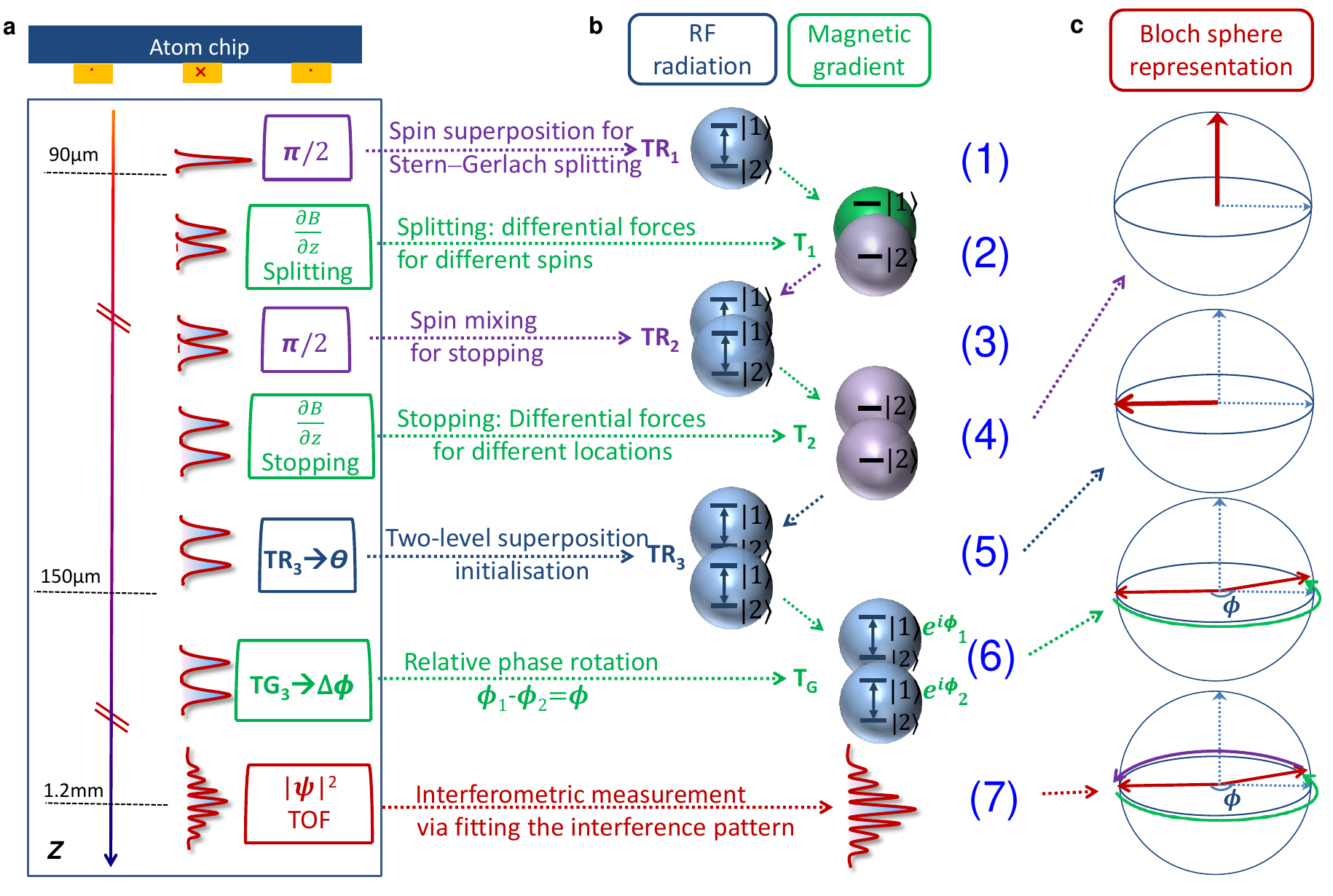}
\caption{\label{fig-SchemeS1} \textbf{Detailed Scheme of the Clock Interferometer.} In {\bf a}, we show the sequential steps in the experiment, depicted schematically\,(not to scale). The experiment takes place in free fall along the z-axis. In {\bf b}, we present the evolution of the states along the sequence. After the atoms are released from the trap, one radio-frequency $\pi/2$ pulse\,($TR_1$) is applied to create an equal superposition of the $|2 \rangle\equiv|F=2,m_F=2\rangle$ and $|1 \rangle\equiv|F=2,m_F=1\rangle$ states\,(step \textbf{1}). These two spin states are then exposed to a differential force created by a magnetic gradient pulse ${\partial B}/{\partial z}$
of duration $T_1$\,(step \textbf{2}),
generated by currents in the atom chip wires, leading to different forces and, as a result, different positions and final velocities of the two states.
A second $\pi/2$ pulse\,($TR_2$)\,(step \textbf{3}) is applied to mix the spins in each one of the wave packets, and then to stop the relative velocity of the wave packets, a second magnetic gradient pulse\,($T_2$)\,(step \textbf{4}) is applied to yield differential forces for the same-spin states at different locations.
As during $T_2$, the $|1 \rangle$ state from the two wave packets are pushed outside the experimental zone, the system then consists of two wave packets in the $|2 \rangle$ state\,
(separated along the z-axis, with zero relative velocity). The internal superposition is then initialised with an RF pulse\,(step \textbf{5}) of duration $TR_3$\,($T_R$ in the text). Subsequently, the relative phase of the wave packets can be adjusted by applying a magnetic field gradient $\partial B/\partial z$ for a duration of $T_G$\,(step \textbf{6}). Finally, the wave packets can expand and overlap before being captured by a charge-coupled device camera. This interference process also acts as a projection operation between the two vectors as represented by the geodesic connection line\,(step \textbf{7}).
In {\bf c}, we show the evolution of the states on the surface of the Bloch sphere during the sequence. After preparing two coherent wave packets at distinct locations\,(step \textbf{(4)}), an RF pulse of duration $TR_3$\,($T_R$ in the text) is applied to shift the two vectors from the north pole to the equatorial plane of the Bloch sphere to initialise the internal superposition\,(step \textbf{5}). A magnetic field gradient of duration $T_G$ induces the relative phase $\phi$\,(step \textbf{6}), and the interferometric measurement is done via fitting the interference pattern\,(step \textbf{7}).}
\end{figure*}

In Fig.\,\ref{fig-SchemeS1}, we present the details of our experimental procedure. This includes the splitting of a single Bose--Einstein condensed\,(BEC) wave packet into a pair in different locations\,(steps 1--3), the halt of their mutual velocity\,(step 4), the initiation of a two-level superposition inside each wave packet\,(step 5), the introduction of a relative phase\,(step 6), and the generation of an interference pattern during the time-of-flight process\,(step 7). Our experiment leverages an atom chip setup as discussed in the literature by \cite{keil2016fifteen}. We start by establishing a BEC comprised of approximately $10^4$ Rubidium-87 atoms in the state $|2\rangle\equiv|F=2,m_F=2\rangle$. This BEC is roughly 90\,$\mu$m beneath the chip surface in a magnetic trap. After releasing the BEC atoms from the trap, the entire experimental sequence occurs in a homogeneous magnetic bias field of 36.7\,gauss in the y-direction\,(with z being the direction of gravity). This field establishes an effective two-level system, with $|1 \rangle\equiv|F=2,m_F=1\rangle$ and $|2 \rangle\equiv|F=2,m_F=2\rangle$, via the non-linear Zeeman effect. Here, $E_{ij}=E_{21}$ is approximately $h\times25\,$MHz\,(where \textit{i},\,\textit{j} are the $m_F$ values, all within the $F=2$ manifold, and $h$ is the Planck constant), and $E_{21}-E_{10}$ is approximately $h\times180$\,kHz.

Next, we introduce a radio-frequency\,(RF) pulse with a typical duration of $10\,\mu$s, resulting in a $\theta=\pi/2$ rotation. This prepares a spin superposition, $(|1\rangle+|2\rangle)/\sqrt{2}$, between the
$|2\rangle$ and $|1 \rangle$ states. A magnetic gradient pulse, represented as ${\partial B}/{\partial z}$ and lasting $T_1=4\,\mu$s, is then applied. This pulse, produced by currents in the atom-chip wires, facilitates a Stern--Gerlach splitting, where spins experience differential forces. To enable interference between the two wave packets\,(given $|2\rangle$ and $|1\rangle$ are orthogonal), a second $\pi/2$ pulse\,($TR_2$) is applied to mix the spins. 
To stop the relative velocity of the wave packets, a subsequent magnetic gradient pulse\,($T_2$) is used. This generates differential forces for the same-spin states positioned dissimilarly. A spatial superposition of two wave packets in state $|2 \rangle$ now exists\,(separated along the z-axis). Note, during $T_2$, the $|1 \rangle$ state from the two wave packets is expelled from the experimental region. 

The $\theta$ control (introduced in Fig.\,\ref{fig1}a) is achieved using a third RF pulse with a duration of $TR_3$\,(referred to as $T_R$ in the main text). The relative rotation, $\phi$, between the wave packets can be adjusted with a third magnetic field gradient, lasting $T_G$. The wave packets then undergo expansion and overlapping to produce the interference pattern, with a time-of-flight of approximately 10 ms\,(significantly exceeding the reciprocal of the trap frequency, roughly 500 Hz). Finally, we capture an image through absorption imaging\,(see Fig.\,\ref{RawFigure}).

Three parallel gold wires on the chip surface produce the magnetic gradient pulses. These wires have a length of 10 mm, a width of 40\,$\mu$m, and a thickness of 2\,$\mu$m. The chip wire current is generated by a simple 12.5 V battery and is modulated using a homemade current shutter. The three parallel gold wires are spaced 100\,$\mu$m apart\,(center to center), with the same current flowing through them in alternating directions. In contrast with using a single gold wire, this three-wire configuration creates a 2D quadrupole field at $z=100\,\mu$m below the atom chip. Given that magnetic-field fluctuations are proportional to the field strength, and the primary source of the instability stems from the gradient pulses\,(since the bias fields from external coils are highly stable), placing the atoms near the center\,(or zero point) of the quadrupole field effectively reduces the magnetic-field noise. This is achieved by the three-wire configuration while preserving the magnetic gradients' strength.
 
\section{Raw interferogram images}

\begin{figure}
\begin{center}
\includegraphics[scale=1]{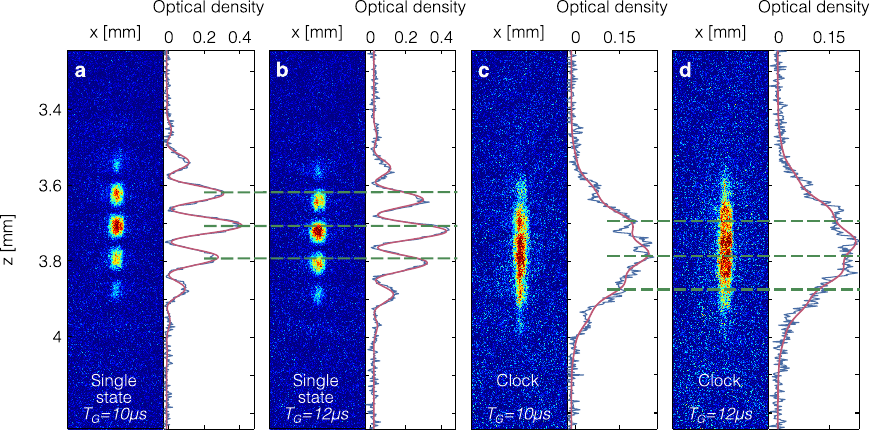}
\end{center}
\caption{\textbf{Raw Interferograms with Fit Functions.} The raw interferogram images are fitted using a sinusoidal function with a Gaussian envelope\,(red curves, see details in SM text). The four single-shot raw interferograms shown are part of the data used in Fig.\,\ref{fig2}\textbf{c}. In \textbf{a} and \textbf{b}, we show the single-state interference when the field gradient pulse duration changes from 10\,$\mu$s to 12\,$\mu$s. In \textbf{c} and \textbf{d}, we also present the clock interference from 10\,$\mu$s to 12\,$\mu$s, namely, before and after the $\pi$ phase change. The $\chi^2$ values are 0.98, 0.97, 0.96, and 0.94, respectively, and the fit phase errors are 0.016, 0.019, 0.14, and 0.17\,rad, respectively. The average error of the sixteen clock interference shots that make up the two data points in Fig.\,\ref{fig2}\textbf{c} is 0.2\,rad. The raw data presented here were gathered at the Atom Chip Lab at Ben-Gurion University of the Negev\,\cite{QuCom, GeoPhaseJump}.}
\label{SingleShot}
\end{figure}

\begin{figure}
\begin{center}
\includegraphics[scale=0.5]{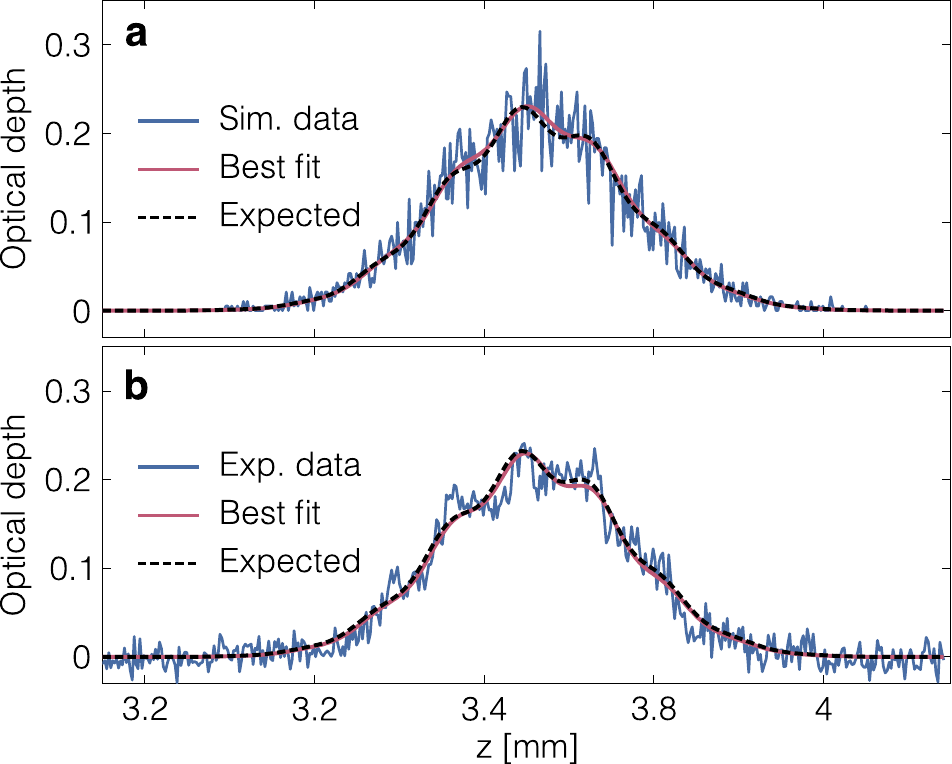}
\end{center}
\caption{\textbf{Monte Carlo Simulation and Raw Data Fit.} In \textbf{a}, we show a comparison of a Monte Carlo simulation where the expected distribution is sampled using five thousand atoms\,(the blue solid curve), which stochastically follow the theoretical distribution\,(the black dashed curve). The expected distribution is given by Eq.\,\eqref{EQ:profile} with the inferred visibility $v$ and phase $\Phi$. Note that the interferogram is not symmetric due to the difference in position between the peak of the Gaussian envelope and the maximum of the interference pattern, which is determined by the phase $\Phi$. The rest of the parameters, such as the interferogram wavelength and the width and position of the Gaussian profile, are extracted from the fit to the experimental data. The red solid curve shows the sinusoidally modulated Gaussian profile fit to the blue data. In \textbf{b}, we show the results from the same procedure but using experimental data of the eight shots constituting the clock interference data at 12\,$\mu$s. The average fit error for $\Phi$ for these experimental conditions is 0.20\,rad. The average fit error in our simulation\,(panel \textbf{a}) is 0.19\,rad. The raw data presented here were gathered at the Atom Chip Lab at Ben-Gurion University of the Negev\,\cite{QuCom, GeoPhaseJump}.}
\label{RawdataFit}
\end{figure}

\begin{figure}
\begin{center}
\includegraphics[width=0.98\textwidth]{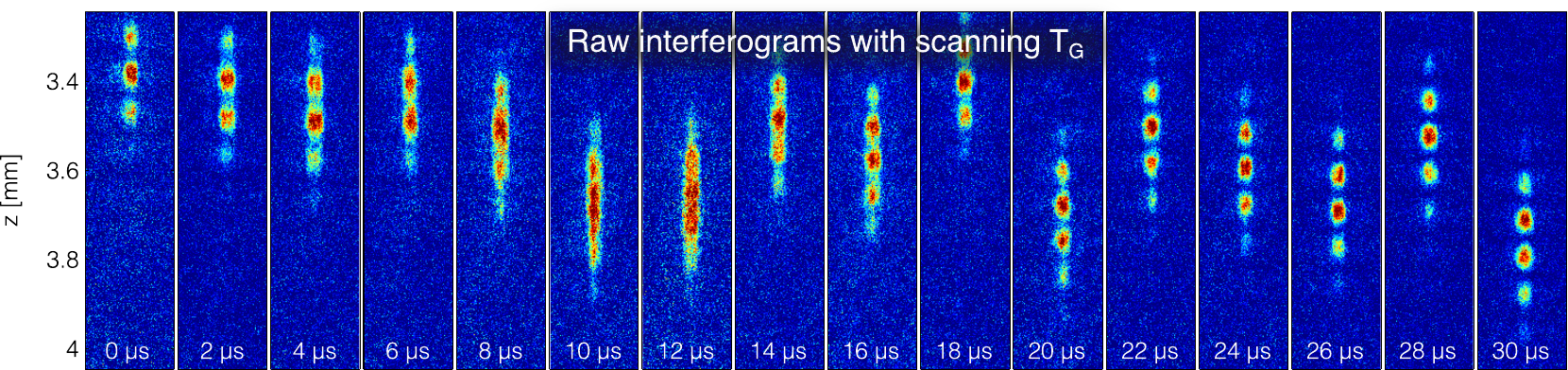}
\end{center}
\caption{\textbf{Raw Interferograms Acquired with Different Field Gradient Durations.} An internal superposition of $P_2/P_1=51.4\%/48.6\%$\,(with $0.4\%$ uncertainty) is prepared. The number in each sub-figure indicates the field gradient duration of $T_G$. The scale for each image is kept consistent to enable direct comparison of the phase. Small inhomogeneities in the bias magnetic field induce an initial relative phase of the clock wave packets, which also affect the single-state interferometer in a similar way\,(see Fig.\,\ref{fig2}\textbf{a}). Due to the longer duration of the $T_G$ pulse, the atoms propagate further downward. The small variations in the interference pattern centre-of-mass location are due to fluctuations in the initial shot-to-shot BEC release conditions in position and momentum, while the inferred phase is stable, as explained in \cite{machluf2013coherent} and \cite{margalit2019analysis}. The raw data presented here were gathered at the Atom Chip Lab at Ben-Gurion University of the Negev\,\cite{QuCom, GeoPhaseJump}.}
\label{RawFigure}
\end{figure}

In Fig.\,\ref{SingleShot}, we show the raw interferogram images corresponding to the phase data presented in Fig.\,\ref{fig2}\textbf{c}, where the four data points\,(two from the single-state interferometer and two from the clock interferometer) are directly compared. The interference pattern is fitted with a sinusoidally modulated Gaussian profile: 
\begin{equation} \label{EQ:profile}
   n(z) = \mathcal{A} \exp\left[-\frac{(z-z_\text{com})^2}{2\sigma_z^2}\right] \left\{1+ v \sin\left[\frac{2\pi}{\lambda}(z-z_\text{ref})+\Phi_T\right]\right\}+c\, ,
\end{equation}
where $\mathcal{A}$ is a constant related to the optical density in the system, $z_\text{com}$ is the centre-of-mass position of the combined wave packet at the time of imaging, $\sigma_z$ is the Gaussian width of the combined wave packet obtained after the time of flight, $\lambda=\frac{ht}{md}$ is the fringe periodicity, $v$ is the visibility, $z_\text{ref}$ is a fixed reference point, $c$ is the background optical density from the absorption-imaging, and $\Phi_T$ is the phase of the interference pattern that appears in Eq.\,\eqref{EQ-Phi}. In the fringe periodicity $\lambda=\frac{ht}{md}$, $h$ is the Planck constant, $t$ is the time-of-flight duration, $m$ is the mass of a $^{87}$Rb atom, and $d$ is the distance between the two wave packets. In Fig.\,\ref{SingleShot}\textbf{a} and Fig.\,\ref{SingleShot}\textbf{b}, $\Phi_T = \phi_2$\,[Eq.\,\eqref{EQ-Phi}] since the entire population is in state $\ket{2}$, where the change of the single-state interference phase is highlighted by the green dashed lines. In Fig.\,\ref{SingleShot}\textbf{c} and Fig.\,\ref{SingleShot}\textbf{d}, the clock interference phase exhibits a phase jump, with peaks in Fig.\,\ref{SingleShot}\textbf{c} corresponding to valleys in Fig.\,\ref{SingleShot}\textbf{d}.

We further study the fitting error in Fig.\,\ref{RawdataFit}. To do so, we analyse the results for the clock interferometer at $T_G = 12$\,$\mu$s\,(the one shown in Fig.\,\ref{SingleShot}\textbf{d}) by performing a Monte Carlo simulation where the expected distribution of 5,000 atoms was sampled, and compared it with the experimental results. We calculated the expected distribution using Eq.\,\eqref{EQ:profile} and the inferred visibility $v=v(\theta, \phi)$ and phase $\Phi(\theta, \phi)$. The rest of the parameters are extracted from the fit to the experimental data. Both cases demonstrate that the fit is able to recover the expected distribution\,(thus $v$ and $\Phi$). We obtain a fit error for $\Phi$ of $0.19$\,rad in the simulation. This is close to the experimental values, which average $0.2$\,rad. Therefore, we conclude that the main source of fit error comes from the fringe visibility of the probability distribution. Matching fit errors are also obtained from a simulation and experiment for the single-state interferometer and are shown in Fig.\,\ref{SingleShot}\textbf{a} and Fig.\,\ref{SingleShot}\textbf{b}. In this procedure, we use a simple fitting method, and we expect the fitting errors to further improve using more advanced fitting methods.

In Fig.\,\ref{RawFigure}, we show the full-range\,(2$\pi$) raw interferogram images corresponding to the clock interference phase data presented in Fig.\,\ref{fig2}. The entire scanning range of $T_G$ is from 0\,$\mu$s to 30\,$\mu$s, reaching one full cycle\,(2$\pi$) of the relative phase rotation.

\section{Noise analysis}
Here, we analyse in more detail the different noise sources, how they impact the measured total phase, and how they scale with respect to the phase slope.

Let us consider that the number of atoms in states $\ket{1}$ and $\ket{2}$ are $N_1$ and $N_2$, respectively, after the RF pulse. Then, $R = \cos^2(\theta/2)/\sin^2(\theta/2) = N_2^2 / N_1^2$, which is the ratio between the populations in the internal degrees of freedom, and $G = [1 - N_2^2 / N_1^2]^{-1}$, which is the phase-change slope. If there is an error in the RF pulse, the population will fluctuate $N_i = N_i^0\,(1 + n_i)$, and then the slope will also deviate from the average value $G^0$ by $\Delta G$, resulting in an extra phase given by
\begin{equation}
\Delta \Phi_\text{amp} = \Delta G \Delta \phi_{\text{sig}} \approx 2\,(G^0)^2\,(n_2 - n_1) \Delta \phi_{\text{sig}}.
\end{equation}
If atoms are lost, they will do so independently of the internal state, so $n_2 \approx n_1$, and consequently, there will not be a differential signal. In addition, infidelity in the pulse that creates the superposition will be amplified with the slope. The effect can be attenuated by preparing the quantum superposition as precisely as possible. In our experiment, we do not observe an increase in noise proportional to $(G^0)^2$. Thus, we are in the case where the RF pulse is precise enough to suppress this noise.

If both wave packets have extra phases due to noise, $\phi_i = \phi_0^i + \Delta \phi^i$ with $i=A, B$, then the error in the total phase will be
\begin{equation}
    \Delta \Phi_T^{\text{phase}} = G\,(\Delta \phi^B - \Delta \phi^A) + \Delta \phi^A \ .
\end{equation}
However, as both wave packets are affected by the same noise sources,  $\Delta \phi^B$ is highly correlated to $\Delta \phi^A$, thus $\Delta \phi^B \approx \Delta \phi^A \approx \Delta \phi$. Consequently, $\Delta \Phi_T^{\text{phase}} = \Delta \phi$, which is not amplified or affected by the visibility. As highlighted in a related study involving an RF interferometer\,\cite{RF2015photonic}, the output phase error remains constant, irrespective of the slope and phase amplification. An analogous analysis showing the high phase coherence of the two wave packets is presented in Ref.\,\cite{margalit2019analysis}. Summing up, the amplification of the phase-change signal is primarily attributed to the geometric phase amplification and the geodesic rule, which do not introduce significant additional noise\,\cite{Vedral2003, GP2003, Rauch2009, BerryExperi2013, yale2016optical}.

In our experiment, the signal $\phi$ results from an external field that our interferometer attempts to estimate has a certain uncertainty $\delta \phi$. This external field is simulated in our experiment by a magnetic gradient of duration $T_G$. The uncertainty in the magnetic gradient comes mainly from fluctuations in the driving current, duration, and wavepacket's positions\,(see the section of detailed experimental procedure in SM). Relative current fluctuations when applying the test field $\delta I/I$ and timing uncertainty $\delta T/T$ contribute to the uncertainty of the phase to be estimated. We estimate a root-mean-square value of $\sim 10^{-3}$ for both of them. As both the current and time are proportional to $\phi$, we find that $\delta \phi/\phi \approx \delta I/I + \delta T/T$. Therefore, we estimate an uncertainty contribution to $\phi$ of approximately $0.2\%$, which is consistent with previous calculations\,(specifically in \cite{machluf2013coherent, margalit2019analysis}). Another source of uncertainty in $\phi$ is the position of the initial wavepacket, which fluctuates by $1\%$. As the magnetic gradient scales inversely proportional to the squared distance, we estimate an uncertainty contribution to $\phi$ of approximately $2\%$ ($2.2\%$ in total). These fluctuations are smaller than the width of the data points in Fig.\,\ref{fig2} and Fig.\,\ref{EstimatedGain} and do not contribute significantly to the precision estimation. 

Finally, our current analysis treats the technical noise as a constant. In a hypothetical scenario in which the technical noise is more complicated, the argument that we described in detail in this work still stands; the influence of the technical noise on the measurement uncertainty of $\phi$ is suppressed because of the amplification, and the quantum noise ultimately limits the total measurement uncertainty.

\section{Sensitivity calculation}
The two red data points highlighted in Fig.\,\ref{fig2}\textbf{c} correspond to -1.36 rad and -5.71 rad over a relative rotation change from 0.94 $\pi$ to 1.04 $\pi$. This results in a phase-changing slope $\partial \Phi / \partial \phi$ of 13.85. The measurement errors $\Delta \Phi$ associated with these values are 0.186 rad and 0.192 rad, respectively. This results in an estimated sensitivity, $\Delta^2\phi$, of $3.73\times10^{-4}$. The two blue points from Fig.\,\ref{fig2}c correspond to -2.93 rad and -3.68 rad over a relative rotation change from 0.94 $\pi$ to 1.04 $\pi$. This results in a 
phase-changing slope $\partial \Phi / \partial \phi$ of 2.39. The measurement errors $\Delta \Phi$ associated with these values are 0.085 rad and 0.094 rad, respectively. This results in an estimated sensitivity, $\Delta^2\phi$, $2.80\times10^{-3}$. The geometric phase measurement yields an 8.8 dB enhancement in sensitivity compared to the conventional dynamical phase measurement. 

\section{Connection to other types of matter--wave interferometers}

To connect our scheme to other types of matter--wave interferometers, we note that in Fig.\,\ref{fig-Scheme}\textbf{a}, both arms are populated with the same spin state, which corresponds to a Bragg interferometer\,\cite{GiltnerPRL1995}. One can envision another scenario in which one arm is in state $\ket{1}$, and the other is in state $\ket{2}$, similar to a Kasevich--Chu interferometer\,\cite{KasevichPRL1991}. The phase at the output of the interferometer will be the difference between the phase accumulated at each arm. As no phase amplification is present, we expect these scenarios to perform similarly to the one in Fig.\,\ref{fig-Scheme}\textbf{a}. Those approaches could also benefit from a geometric phase amplification if an internal superposition is prepared during the interrogation stage, as shown in our approach in Fig.\,\ref{fig-Scheme}\textbf{b}. In addition, it should be mentioned that experimentally acquired interferograms contain additional information we have not exploited in this study, for example, the interferometric fringe visibility. In this case, there is no need to avoid $\theta=\pi/2$. However, the phase amplification effect we intended to discuss here is specific to the interferometric phase $\Phi_T$.

\section{Geometric phase and geodesic rule for different internal superposition}

\begin{figure*}[h!]    
\centering
\includegraphics[scale=0.5]{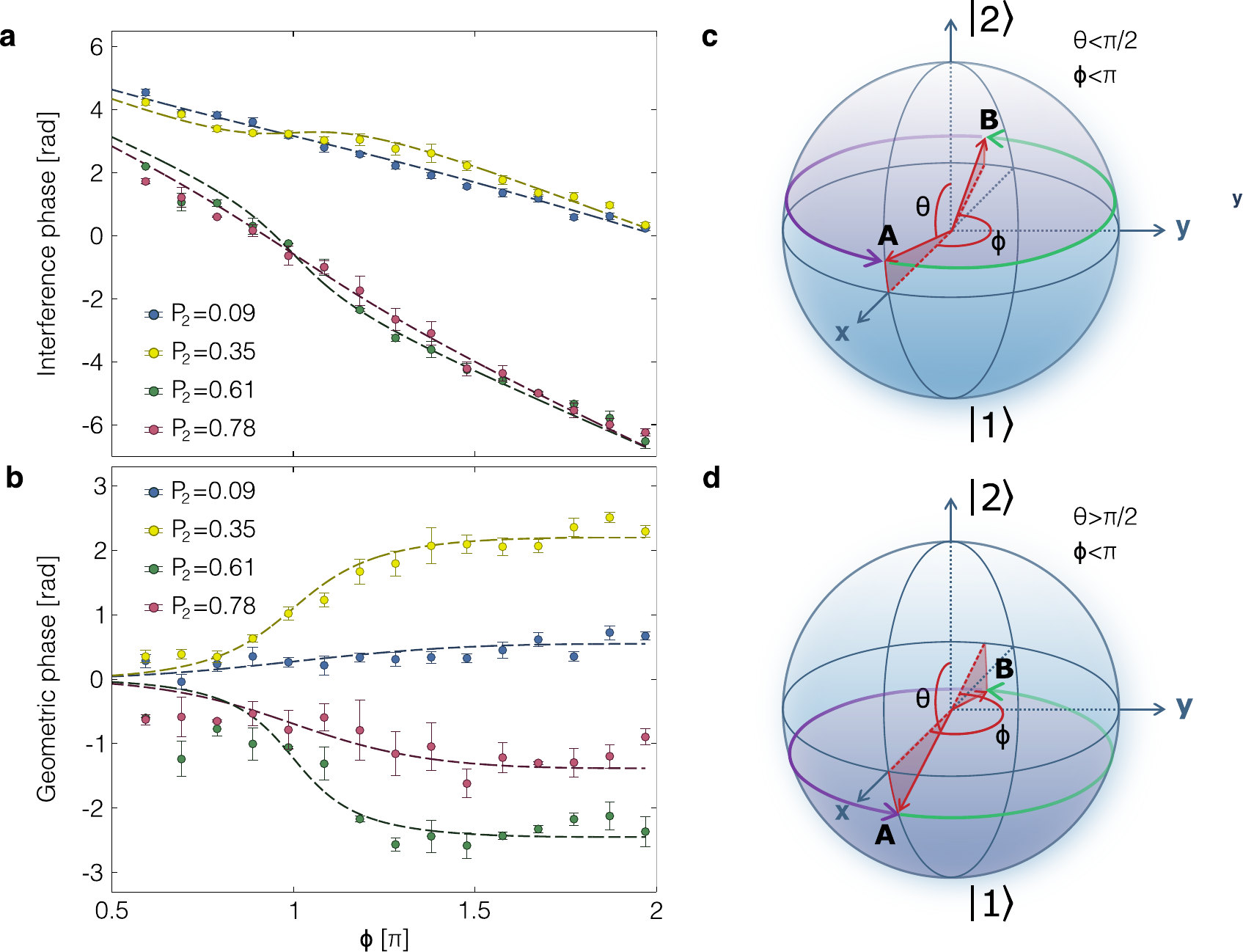}
\caption{\textbf{Geometric Phase and Geodesic Rule for Different Internal Superposition.} In {\bf a}, we measure total phase $\Phi_T$ as a function of $T_G$\,($\phi$) for the population of $m_F=2$; $P_2$, is equal to 0.09, 0.35, 0.61 and 0.78, respectively. The dashed lines are the result of Eq.\,\eqref{EQ-Phi} with the corresponding populations in state $m_F=2$. Each data point is an average of eight experimental cycles, and the error bars are the standard error of the mean\,(SEM) in this subsample. In \textbf{b}, we remove the DP\,(estimated analytically by the values of $m_F=2$) and keep only the GP, to further confirm the geodesic rule and the connection between the enclosed area and the GP. The GP becomes significant when the system approaches an equal superposition of internal states and flips its sign when $P_2$ is larger than 0.5. In {\bf c-d}, we illustrate the geometric nature of the results. The geodesic rule dictates that the GP is larger when Bloch vectors \textbf{A} and \textbf{B} approach the equator of the Bloch sphere. It also explains the sign flip as the enclosed area moves from the north to the south hemisphere. The raw data presented here were gathered at the Atom Chip Lab at Ben-Gurion University of the Negev\,\cite{QuCom, GeoPhaseJump}.}
\label{fig4}
\end{figure*}

As discussed in the main text, the total phase has two components, the geometric and dynamical phases, and both have a certain tendency with respect to $\phi$. As the sensitivity increases with the slope of the total phase, it would be desirable that in the working point $\phi \approx \pi$, the slope sign of the geometric and dynamical phases is the same, thus maximising sensitivity. In Fig.\,\ref{fig4}\textbf{a}, we illustrate this point by showing the measured interference phase, also known as the total phase, obtained for different population levels in the state $\ket{2}$. The population levels considered were $0.09$, $0.35$, $0.61$, and $0.78$. We observe a close agreement between the measured and predicted values when fitting the data to Eq.\,\eqref{EQ-Phi}. When the system is prepared close to a single state\,($P_2=0.09$ and $P_2=0.78$), the GP diminishes and tends towards zero\,(Fig.\,\ref{fig4}\textbf{b}). In such cases, the interferometric case approaches a linear function. When the system is prepared closer to a balanced superposition\,($P_2 = 0.35$ and $P_2=0.61$), the prominence of the GPs is evident. Due to the geodesic rule\,\cite{samuel-bhandari, bhandari19912, GeoPhaseJump}, the GP sign is closely connected with the enclosed area on the surface of the Bloch sphere and its normal direction\,(Fig.\,\ref{fig4}\textbf{c} and Fig.\,\ref{fig4}\textbf{d}). The GP flips the sign when the population transfer changes from less than 0.5 to larger than 0.5. As a result, one of these phases shares the same sign with the DP, which boosts the phase sensitivity, while the other opposes it, suppressing the phase sensitivity. From a metrological perspective, understanding this behavior is crucial for effectively combining the geometric and dynamical phases. 

\end{document}